\documentclass[aps,prb,reprint,floatfix,superscriptaddress]{revtex4-1}
\usepackage[english]{babel}                                                           
\usepackage[utf8]{inputenc}                                                            
\usepackage[T1]{fontenc}
\usepackage{epsfig}
\usepackage{amsmath}
\usepackage{hyphenat}
\usepackage{hyperref}
\usepackage{placeins}
\usepackage[justification=raggedright]{caption}

\begin{document}

\title{Modification of the Lifshitz-Kosevich formula for anomalous quantum oscillations in inverted insulators}
\author{Simonas Grubinskas}
\affiliation{Institute Lorentz $\Delta$ITP, Leiden University, PO Box 9506, Leiden 2300 RA, The Netherlands}
\affiliation{Institute for Theoretical Physics and Center for Extreme Matter and Emergent Phenomena,
Utrecht University, Leuvenlaan 4, 3584 CE Utrecht, The Netherlands}
\author{Lars Fritz}
\affiliation{Institute for Theoretical Physics and Center for Extreme Matter and Emergent Phenomena,
Utrecht University, Leuvenlaan 4, 3584 CE Utrecht, The Netherlands}

\begin{abstract}
It is generally believed that quantum oscillations are a hallmark of a Fermi surface and the oscillations constitute the ringing of it. Recently, it was understood that in order to have well defined quantum oscillations you do not only not need well defined quasiparticles, but also the presence of a Fermi surface is unnecessary. In this paper we investigate such a situation for an inverted insulator from a analytical point of view. Even in the insulating phase clear signatures of quantum oscillations are observable and we give a fully analytical formula for the strongly modified Lifshitz-Kosevich amplitude which applies in the clean as well as the disordered case at finite temperatures.  
\end{abstract}
\maketitle

\section{Introduction}
\begin{figure}
\includegraphics[width=0.4\textwidth]{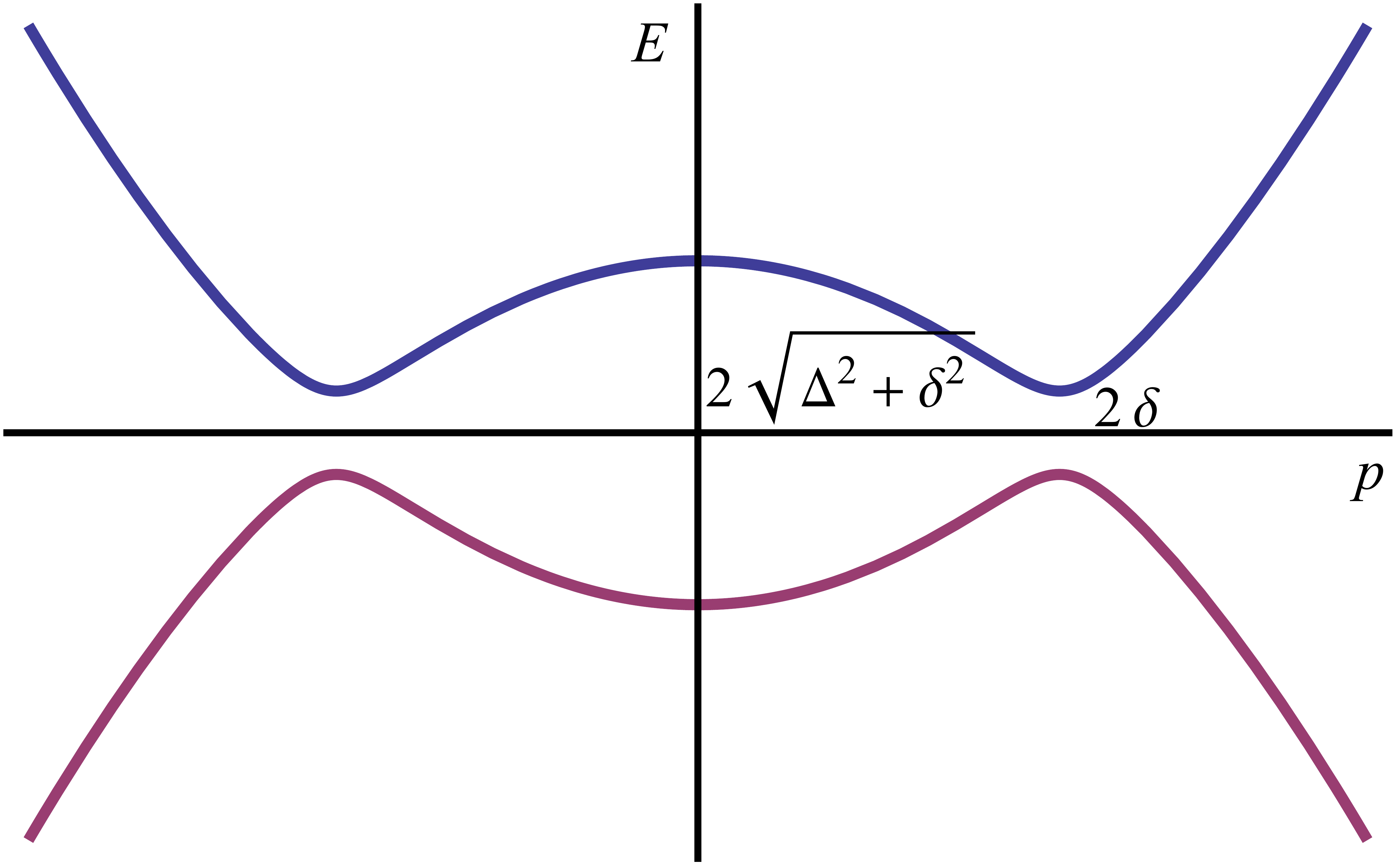}
\caption{Dispersion relation along a one-dimensional cut in a two-dimensional momentum space. }\label{fig:Fig1}
\end{figure}

Landau quantization of electrons in magnetic fields is at the heart of many interesting phenomena: besides being responsible for integer and fractional quantum Hall effects it also plays an important role in determining the electronic structure of (correlated) metallic states. Quantum oscillation measurements provide the basis for understanding electronic properties of metals: they reveal information about the Fermi surface as well as about the quasiparticle effective masses and disorder levels via the Dingle temperature. 
In the standard experimental setup of quantum oscillations, both transport (Shubnikov-de Haas) and thermodynamic (de Haas-van Alphen) quantities are measured as functions of the inverse magnetic field and display periodic behavior whose period is set by the cross section of the Fermi surface encircled by the electrons in a semiclassical picture. In uncorrelated metals, the amplitude of the oscillations is described by the Lifshitz-Kosevich (LK) formula~\cite{LifshitzKosevich,Luttinger1960,Shoenberg1984}. This formula describes the damping of the oscillations due to thermal, disorder, and interaction broadening. There are few cases in which deviations from the standard LK form have been shown theoretically~\cite{Mirlin2006,Martin2003,FowlerPrange,EngelsbergSimpson,Kueppersbusch,Knolle2015}.
More recently, it was shown that inverted insulators (Fig.~\ref{fig:Fig1}) can show well defined anomalous quantum oscillations even in the absence of a Fermi surface. In the meantime several authors have discussed this situation for different model systems~\cite{Falkovsky,Knolle2015,Knolle2016,Zhang2016,Alisultanov2016,Pal2016,Ram2017}. 

The main result of this work compared to previous works is that we present a fully analytical version of the LK formula which allows for very simple approximate solutions in all conceivable limits, including weak potential disorder.

The paper is organized as follows. We start in Sec.~\ref{sec:model} by introducing a model system which describes an inverted insulator. A possible realization of this could be bilayer graphene in a perpendicular electric field. We concentrate on the Landau level structure and discuss its semicalssical limit. We then move to a calculation of the grand potential in Sec.~\ref{sec:grandpot} where we follow the treatment of Luttinger~\cite{Luttinger1960}. We finish with a discussion of the different regimes in Sec.~\ref{sec:disc} and a conclusion (Sec.~\ref{sec:conclusion}).

\section{Model and Landau level structure}\label{sec:model}
We consider a generic two-band model of the type~\cite{McCann2006}
\begin{eqnarray}
H=\frac{1}{2\tilde{m}}\left(\begin{array}{cc}
2\tilde{m}\tilde{\Delta}- |p|^2  & vp^{*}p^{*}\\
vpp & -2 \tilde{m} \tilde{\Delta}+ |p|^2 
\end{array}\right),
\end{eqnarray}
where momentum $p=p_{x}+ip_{y}$  and $p^*$ denotes its complex conjugate and the parameter $\tilde{\Delta}$ describes the electric field. 
For the dimensionless parameter $v=0$, the spectrum consists of two parabolas that
cross each other (for $\tilde{\Delta}>0$). For $v\ne0$ a gap is
opened and the following relabeling is used to fix the behavior of
the system for large momenta, $\tilde{\Delta}=\Delta\sqrt{1+v^{2}}$
and $\tilde{m}=m\sqrt{1+v^{2}}$. We also introduce $\delta=v \Delta$. The spectrum of this Hamiltonian is given by
\begin{eqnarray}
E=\pm\sqrt{\left(\frac{p^{2}}{2m}-\Delta\right)^{2}+\delta^{2}}\;,
\end{eqnarray}

from which we see that $\delta$ measures the gap of the system.

\subsection{General Landau level structure}

A perpendicular magnetic field is introduced via minimal coupling and we subsequently specify the Landau gauge, {\it{i.e.}}, $\vec{A}=B(0,x,0)^T$. We make an ansatz for the wavefunction of the type $\Phi (x,y)=e^{i k_y y} \phi_{k_y}(x)$ and introduce the raising and lowering operators $a=\sqrt{\frac{m \omega_c}{2}}(x+\frac{i}{m \omega_c}p_x)$ and $a^\dagger=\sqrt{\frac{m \omega_c}{2}}(x-\frac{i}{m \omega_c}p_x)$ ($[a,a^\dagger]=1$) where $\omega_c=\frac{eB}{m}$ is the cyclotron frequency. This constitutes a basis change and allows us to rewrite the Hamiltonian as
\begin{eqnarray}
H'=\left(\begin{array}{cc}
\tilde{\Delta}-\omega_{c}\frac{a^{\dagger}a+1/2}{\sqrt{1+v^{2}}} & -\frac{v}{\sqrt{1+v^{2}}}\omega_{c}a^{\dagger}a^{\dagger}\\
-\frac{v}{\sqrt{1+v^{2}}}\omega_{c}aa & -\tilde{\Delta}+\omega_{c}\frac{a^{\dagger}a+1/2}{\sqrt{1+v^{2}}}
\end{array}\right)\;.
\end{eqnarray}
The operators $a$ and $a^\dagger$ act as raising and lowering operators on the harmonic oscillator eigenstates defined by $a |N \rangle=\sqrt{N}|N-1\rangle$ and $a^\dagger |N \rangle=\sqrt{N+1}|N+1\rangle$.

Employing a further ansatz $\phi_{k_y}(x+k_y/eB)$, where $\phi_{k_y}(x)=\left(\begin{array}{c}\alpha\langle x|N\rangle\\ \beta \langle x|N-2\rangle \end{array}\right)$
we reduce the problem to a $2\times 2$ Hamiltonian for a two-component spinor $(\alpha,\,\beta)^T$, given by
\begin{eqnarray}
H''=\left(\begin{array}{cc}
\tilde{\Delta}-\omega_{c}\frac{N+1/2}{\sqrt{1+v^{2}}} & -\frac{v\omega_{c}}{\sqrt{1+v^{2}}}\sqrt{N(N-1)}\\
-\frac{v\omega_{c}}{\sqrt{1+v^{2}}}\sqrt{N(N-1)} & -\tilde{\Delta}+\omega_{c}\frac{N-3/2}{\sqrt{1+v^{2}}}
\end{array}\right)\;.\nonumber \\
\end{eqnarray}
In the semiclassical regime of quantum oscillations, {\it i.e.}, $N\gg1$ the Hamiltonian becomes
\begin{eqnarray}
H''\approx\left(\begin{array}{cc}
\tilde{\Delta}-\omega_{c}\frac{N}{\sqrt{1+v^{2}}} & -\omega_{c}\frac{v}{\sqrt{1+v^{2}}}N\\
-\omega_{c}\frac{v}{\sqrt{1+v^{2}}}N & -\tilde{\Delta}+\omega_{c}\frac{N}{\sqrt{1+v^{2}}}
\end{array}\right)\;.
\end{eqnarray}
The energies of the quantized Landau levels that follow from this Hamiltonian are
\begin{eqnarray}
E_{N}=\pm\sqrt{\left(\omega_{c}N-\Delta\right)^{2}+\delta^{2}}\;.\label{eq:symmspectrum}
\end{eqnarray}
The effective cyclotron frequency $\omega_{c}$ describes the general behavior of the system under
the influence of the magnetic field, that is, $\omega_{c}\sim B$. 

If the gap $\delta=0$, the quantity $2\pi m\Delta$ measures the area enclosed
by the Fermi surface in $k$-space. For any finite $\delta>\mu$ the
area remains the same and is equal to the area enclosed by the circular
minimum of the band (we call it `shadow' Fermi surface). As is shown in the next section, this area gives
the period of the quantum oscillations whether we have a real Fermi surface or not.

\section{Grand potential and modified Lifshitz-Kosevich formula}\label{sec:grandpot}

\subsection{Clean case}

The grand potential in a clean non-interacting system is given by
\begin{eqnarray}\label{eq:grandpotential}
\Omega=-T\; {\rm{tr}} \ln (-\hat{G}^{-1})
\end{eqnarray}
where $\hat{G}$ is the Green function of the non-interacting system and $T$ is the temperature. Using the spectrum defined in Eq.~\eqref{eq:symmspectrum} we can write the grand potential as
\begin{eqnarray}
\Omega &=&-D T \sum_{\omega_n} \sum_{N=0}^\infty \sum_{\lambda=\pm} \;  \ln \left(i \omega_n+\mu+\lambda E_N \right) \nonumber \\ &=& -D T \sum_{\omega_n} \sum_{N=0}^\infty  \;  \ln \left( (i \omega_n+\mu)^2- E_N^2 \right)
\end{eqnarray}
with $\omega_n=(2n+1)\pi T$ and the Landau level degeneracy factor $D=c\frac{e B L^2}{2 \pi}$, where $c$ counts the number of different species of electrons: spin, valley, etc. Note that we neglect a possible Zeeman effect.

Using the Poisson resummation technique we can decompose the sum over the Landau levels into a sum of integrals which is organized in terms of harmonic oscillations, {\it i.e.}, $\sum_{N=0}^{\infty}f_{N}=\frac{1}{2}f_0+\int_{0}^{\infty}dxf\left(x\right)+2\sum_{l=1}^{\infty}\int_{0}^{\infty}dxf\left(x\right)\cos\left(2\pi lx\right)$, 
where $f(x)$ is the function obtained by replacing the integer parameter $N$ by a continuous real variable $x$.
In metals this allows us to perform semiclassical approximations with the control parameter $T/\omega_c$ (in the clean and non-interacting limit) leading to the famous Lifshitz-Kosevich formula~\cite{LifshitzKosevich,Shoenberg1984}.
Most importantly, we can identify the oscillatory part of Eq.~\eqref{eq:grandpotential} as
\begin{eqnarray}
\tilde{\Omega}&=&2DT\sum_{\omega_n} \sum_{l=1}^\infty \tilde{\Omega}_{l,\omega_n} \;, \nonumber \\
\tilde{\Omega}_{l,\omega_n} &=&-\int_{0}^{\infty}dx \ln \left( (i \omega_n+\mu)^2- E(x)^2 \right) \cos\left(2\pi lx\right) \;. \nonumber \\
\end{eqnarray}
Integrating by parts and ignoring the non-oscillatory boundary terms containing, for instance, Landau diamagnetism we obtain
\begin{eqnarray}
\tilde{\Omega}_{l,\omega_n}=\int_{0}^{\infty}dx \frac{\sin\left(2\pi lx\right)}{2\pi l} \frac{d}{dx}\ln \left( (i \omega_n+\mu)^2- E(x)^2 \right) \;.\nonumber \\
\end{eqnarray}
This can be rewritten as
\begin{eqnarray}\label{eq:integral}
\tilde{\Omega}_{l,\omega_n}= \int_{0}^{\infty}dx \frac{\sin\left(2\pi lx\right)}{2\pi l}\left(\frac{1}{x-x_1}+\frac{1}{x-x_2} \right),
\end{eqnarray}
where $x_{1}=\frac{1}{\omega_{c}}(\Delta+x_{-}+isx_{+})$ and $x_{2}=\frac{1}{\omega_{c}}(\Delta-x_{-}-isx_{+})$
are the roots of the inverse Green function and where $s=\mathrm{sign}(\omega_{n}\mu)$
and
\begin{eqnarray}\label{eq:xpxm}
x_{+}&=& \sqrt{\frac{\sqrt{(\omega_n^2+\delta^2-\mu^2)^2+4 \mu^2 \omega_n^2}+(\omega_n^2+\delta^2-\mu^2)}{2}} \nonumber \\ x_{-}&=& \sqrt{\frac{\sqrt{(\omega_n^2+\delta^2-\mu^2)^2+4 \mu^2 \omega_n^2}-(\omega_n^2+\delta^2-\mu^2)}{2}}\;.\nonumber \\
\end{eqnarray}
For the remainder of the paper we assume that $\Delta$ is the largest energy scale in the problem. A consequence of this is that the poles of Eq.~\eqref{eq:integral} are sufficiently
far into the right half of the complex plane if $\omega_n \ll \Delta$. This allows us to replace $\int_0^\infty dx \to \int_{-\infty}^\infty dx$ in Eq.~\eqref{eq:integral} as long as the Matsubara frequency $\omega_n$ is not too large. However, it can be proven that large $\omega_n$ contributions are exponentially suppressed and the approximation is valid. Extending the integration range over the whole real axis allows us to use the theory of residues for the integral in Eq.~\eqref{eq:integral} leading to

\begin{eqnarray}\label{eq:oscpot}
\tilde{\Omega}&=&2D T \sum_{\lambda=\pm}\sum_{l=1}^{\infty}\frac{1}{l} \sum_{\omega_n>0}e^{-\frac{2 \pi l}{\omega_c} x_+}  \cos\left(\frac{2\pi l}{\omega_c}\left( \Delta+\lambda x_{-}\right)\right)\; \nonumber \\
&=&4D T \sum_{l=1}^{\infty}\frac{1}{l}\cos\left(\frac{2\pi l}{\omega_c} \Delta\right) \sum_{\omega_n>0}e^{-\frac{2 \pi l}{\omega_c} x_+}  \cos\left(\frac{2\pi l}{\omega_c}  x_{-}\right)\;. \nonumber \\
\end{eqnarray}

We observe that the period of the quantum oscillations is governed by the area enclosed by the above-mentioned `shadow' Fermi surface and controlled by the parameter $\Delta$.
Since $\Delta$ is the largest energy scale, we can treat a part of the expression as an amplitude function that modulates quantum oscillations. Let us rewrite Eq.~\eqref{eq:oscpot} as

\begin{eqnarray}\label{eq:oscpotm}
\tilde{\Omega}&=4D \sum_{l=1}^{\infty}\frac{1}{l}\tilde{\Omega}_m\cos\left(\frac{2\pi l}{\omega_c} \Delta\right),  
\end{eqnarray}

where we introduce the quantity

\begin{eqnarray}\label{eq:oscpotmod}
\tilde{\Omega}_m=T \sum_{\omega_n>0}e^{-\frac{2 \pi l}{\omega_c} x_+}  \cos\left(\frac{2\pi l}{\omega_c}  x_{-}\right)\;
\end{eqnarray}
which we call `amplitude' function from here on for reasons to become clear below.

We note that the oscillation frequency is dependent on both the chemical potential and the temperature. 
The general character of the amplitude function $\tilde{\Omega}_{m}$
is that if $x_{+}>x_{-}$ it is mainly an exponential decay function
with slight oscillations, which requires $\pi^{2}T^{2}+\delta^{2}>\mu^{2}$.
On the other hand if $\pi^{2}T^{2}+\delta^{2}<\mu^{2}$ (and so $x_{+}<x_{-}$), 
the amplitude is mainly an oscillatory function.

\subsection{Dilute disorder}

To incorporate disorder in our treatment we modify the non-interacting Green function $-\hat{G}^{-1}$ that enters Eq.~\eqref{eq:grandpotential}. We use the  self-consistent Born approximation for the Green function \cite{Altland} which implies that the grand potential becomes

\begin{eqnarray}\label{eq:grandpotentialDis}
\Omega &=& -D T \sum_{\omega_n} \sum_{N=0}^\infty  \;  \ln \left( (i \omega_n+\frac{i {\rm{sign}}(\omega_n)}{2\tau }+\mu)^2- E_N^2 \right)\;. \nonumber \\
\end{eqnarray}

The calculation presented in the previous subsection proceeds as before and the final result given in equations \eqref{eq:xpxm} and \eqref{eq:oscpot} remains correct but with the replacement

\begin{eqnarray}\label{eq:grandpotentialDisAnw}
\omega_n \rightarrow \omega_n + \frac{{\rm{sign}}(\omega_n)}{2\tau}\;.
\end{eqnarray}

We note, however, that this is not a simple renormalization of the temperature because of the $n$ dependence, meaning we also cannot simply identify a Dingle temperature.

\section{Discussion}\label{sec:disc}
\subsection*{Recovery of Lifshitz-Kosevich formula}

The only case where we can perform the sum over $n$ analytically in \eqref{eq:oscpotmod} is when we can approximate $\omega_{n}^{2}-\mu^{2}+\delta^{2}\approx\omega_{n}^{2}-\mu^{2}$.  This implies that $x_{+}=\omega_{n}$ and $x_{-}=\mu$. In these cases the amplitude reads

\begin{eqnarray}\label{eq:LK-amp}
\tilde{\Omega}_m^{LK} &=& T  \frac{1}{ \sinh \left( \frac{2 \pi^2 l T}{\omega_c} \right)}  \cos\left(\frac{2\pi l}{\omega_c} \mu\right)\;,
\end{eqnarray}

and the oscillatory potential is of the standard Lifshitz-Kosevich type~\cite{LifshitzKosevich,Shoenberg1984},
\begin{eqnarray}\label{eq:LK-reg}
\tilde{\Omega}^{LK} &=&D T \sum_{\lambda=\pm}\sum_{l=1}^{\infty} \frac{1}{l \sinh \left( \frac{2 \pi^2 l T}{\omega_c} \right)}  \cos\left(\frac{2\pi l}{\omega_c}\left( \Delta+\lambda |\mu|\right)\right)\;. \nonumber \\
\end{eqnarray}
This is essentially the result of a metal with two Fermi surfaces cut out of the spectrum of our system by a constant energy level surface $E=\mu$.

There are three situations in which the above approximation holds: (I) the gap is zero ($\delta=0$), so the system has a pair of coinciding Fermi surfaces, one for the electrons and one for the holes; (II) the temperature is much larger than the gap, ($T\gg\delta$), so the temperature blurs the effect of the finite gap; (III) the chemical potential is much larger than the gap ($\mu\gg\delta$), so, again, the gap is thermodynamically insignificant.

\subsection*{Analysis of the amplitude function $\tilde{\Omega}_m$, clean case}

To analyze the behavior of the amplitude $\tilde{\Omega}_m$ we introduce $y=2\pi l/\omega_{c}$. The expression $\overline{\Omega}_{m}=\pi \tilde{\Omega}_{m}y$ then depends on three dimensionless variables, $\overline{T}=\pi Ty$, $\overline{\mu}=\mu y$ and $\overline{\delta}=\delta y$. We now explore the behavior of this function, $\overline{\Omega}_{m}=\overline{\Omega}_{m}\left(\overline{T},\overline{\delta},\overline{\mu}\right)$.

\begin{figure}
\includegraphics[width=0.4\textwidth]{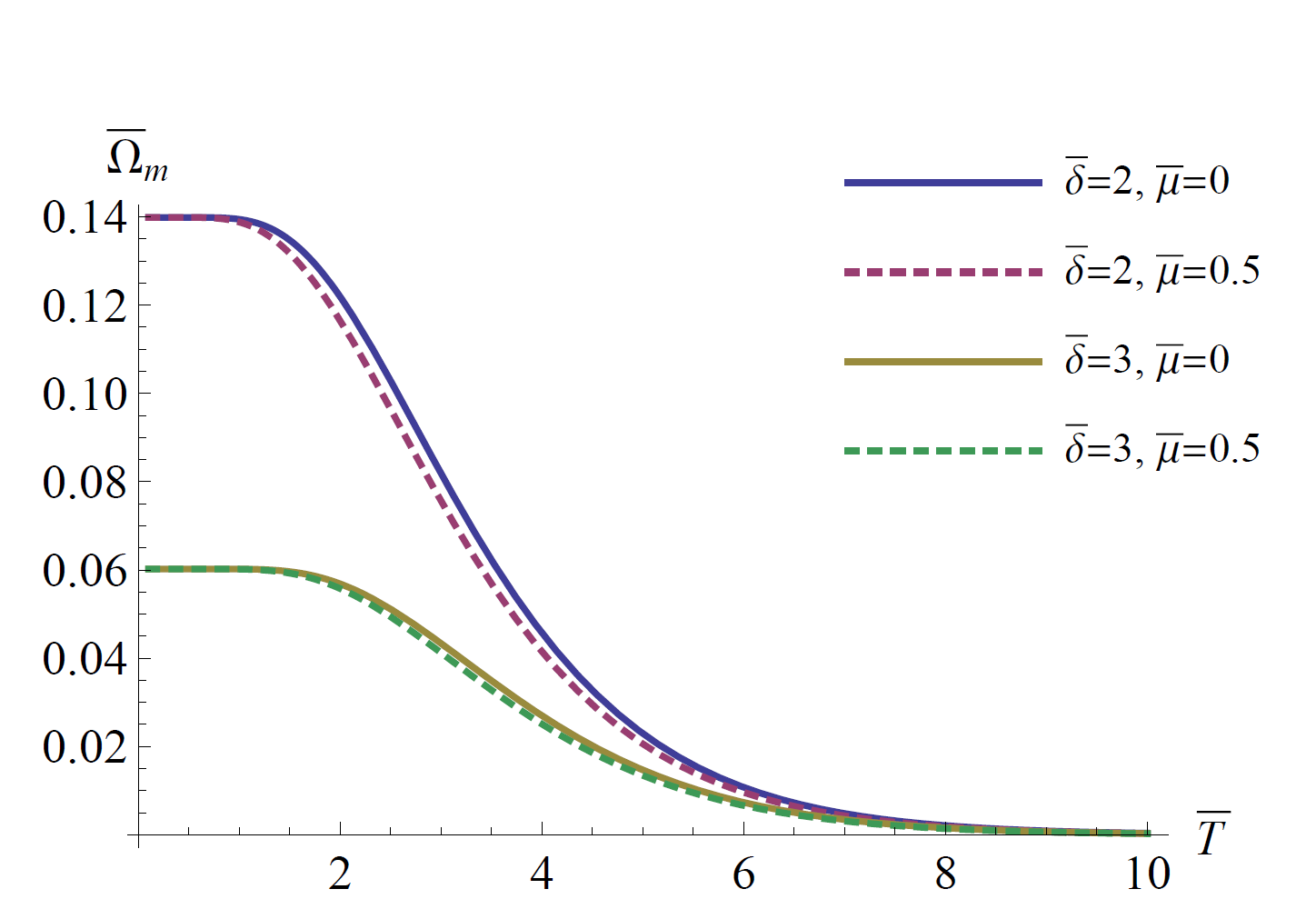}
\caption{The amplitude function $\overline{\Omega}_m$ as a function of the dimensionless temperature $\overline{T}$. We observe that the curves decay exponentially for $\overline{T}>\overline{\delta}$ and flatten out for $\overline{T}<\overline{\delta}$. Also, finite chemical potential $\overline{\mu}$  modifies the result only slightly. }\label{fig:tempdamp}
\end{figure}

\begin{figure}
\includegraphics[width=0.4\textwidth]{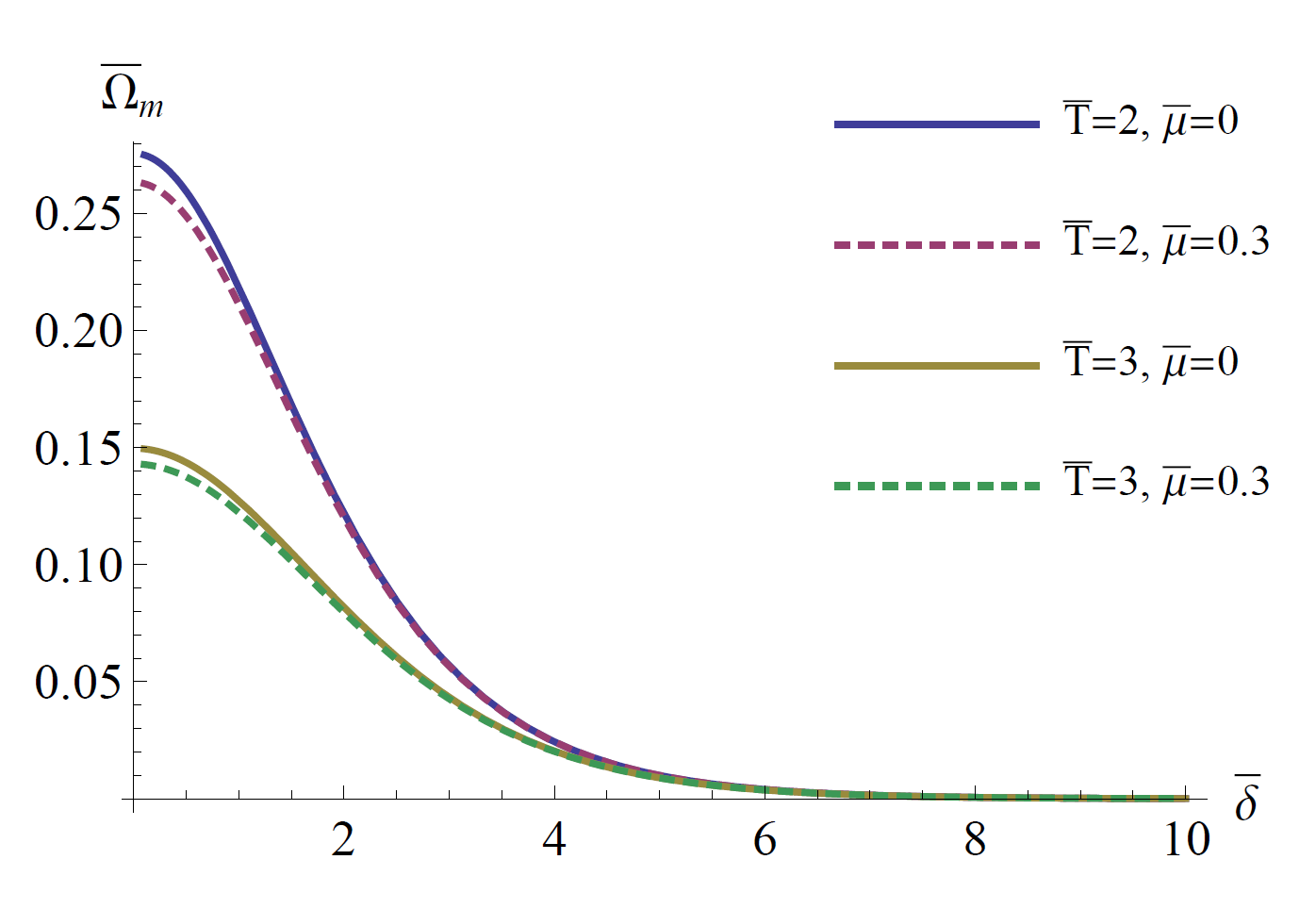}
\caption{The amplitude function $\overline{\Omega}_m$ as a function of the dimensionless gap $\overline{\delta}$. We clearly see that for various values of temperature whenever $  \overline{\delta}>\overline{T} $ the amplitude converges to the same curve characterized by the gap only (that part of the curve also does not depend on the chemical potential as long as $\overline{\mu}<\overline{\delta}$).}\label{fig:gapdamp}
\end{figure}

\begin{figure}
\includegraphics[width=0.4\textwidth]{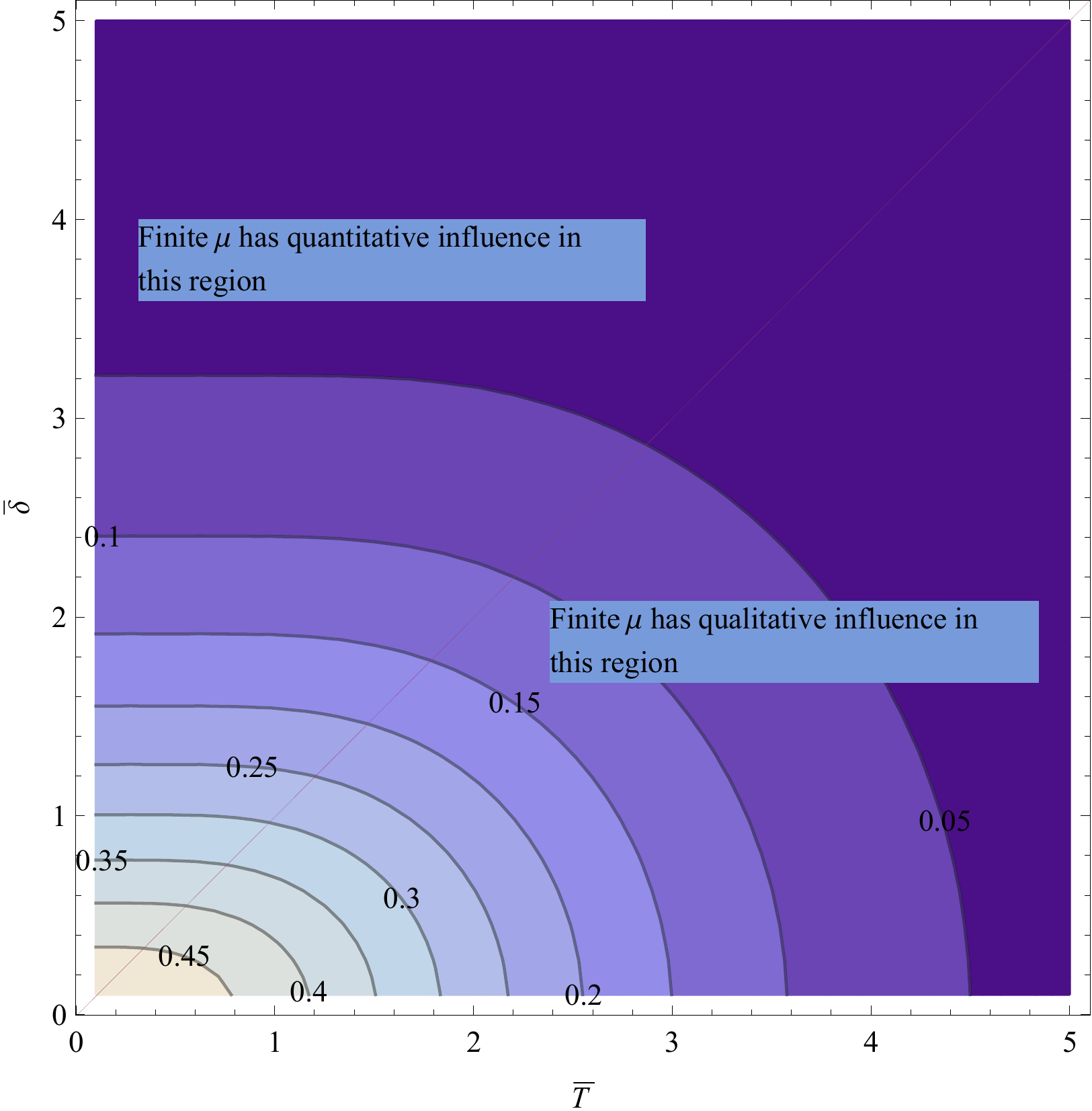}
\caption{Contour plot of the amplitude $\overline{\Omega}_m$ as a function of dimensionless gap $\overline{\delta}$ and dimensionless temperature $\overline{T}$ for $\overline{\mu}=0$. If $\overline{\mu}\ne 0$, the upper side of the graph would not be modified much, whereas the lower side of the graph would be.}\label{fig:contourdamp}
\end{figure}

We first consider the case  $\overline{\mu}<\overline{\delta}$. In general, the amplitude is damped exponentially with increasing  $\overline{T}$ (see Eq.~\eqref{eq:LK-reg} for a specific example). However, from Fig.~\ref{fig:tempdamp} we observe that given that the system is gapped and $ \overline{T}< \overline{\delta}$, the amplitude remains constant, instead of decreasing exponentially. The gap also damps the amplitude because the values of the amplitude for $\overline{T}=0$ do not coincide. We demonstrate how the amplitude is damped by the gap in Fig.~\ref{fig:gapdamp}. In Fig.~\ref{fig:contourdamp} we show a contour plot of the amplitude function for $\overline{\mu}=0$.

If we consider $\overline{T}=0$, we can obtain an exact expression for $\overline{\Omega}_m$ (given $\overline{\mu}<\overline{\delta}$),

\begin{eqnarray}\label{eq:asympt}
\overline{\Omega}_m =\frac{\pi l\delta}{\omega_{c}}K_{1}\left(\frac{2\pi l\delta}{\omega_{c}}\right),
\end{eqnarray} 
where $K_1(x)$ is the modified Bessel function of the second kind.

We can now draw conclusions about the amplitude $\tilde{\Omega}_m$ as a function of inverse magnetic field $y$. If we choose specific $\pi T$ and $\delta$ the amplitude function $\overline{\Omega}_m$ takes values  along the line $\overline{\delta}=k \overline{T}$ in  Fig.~\ref{fig:contourdamp}, where $k=\delta/(\pi T)$.

\begin{figure}
\includegraphics[width=0.4\textwidth]{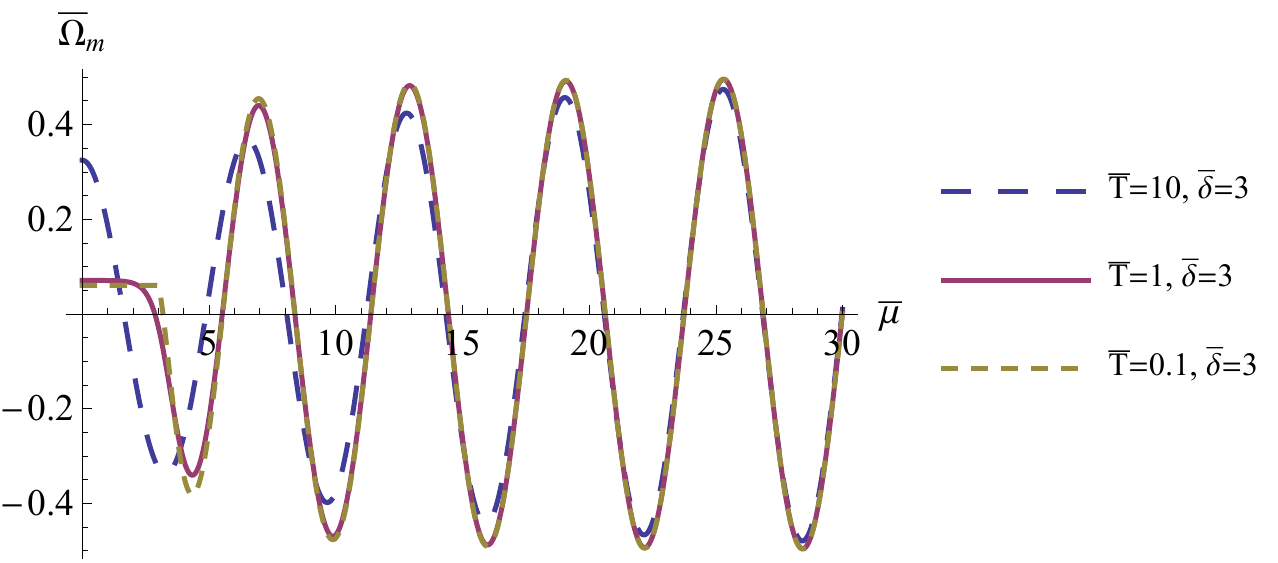}
\caption{Plots of $\overline{\Omega}_{m}$ for different values of $\overline{T}$ but fixed $\overline{\delta}$. We have adjusted the scale to visualize all three functions in one plot because of the severe
damping in the large-temperature case. We observe that all three functions
oscillate with a period $2\pi$ for large $\overline{\mu}$, however
as $\overline{\mu}$ hits the gap $\overline{\delta}$, the function
either keeps oscillating (high temperature $\overline{T}$ case) or
goes to a constant (low temperature case).}\label{fig:mudep}
\end{figure}

Now we consider the case $\overline{\mu}>\overline{\delta}$. In principle, in this case the behavior is  described by  Eq.~\eqref{eq:LK-reg}. We show in Fig.~\ref{fig:mudep} how the amplitude oscillates in the region
$\overline{\mu}>\overline{\delta}$ and is suppressed as soon as  $\overline{\mu}<\overline{\delta}$ (unless $\overline{T}>\overline{\delta}$).
In general it is difficult to describe the behavior when the chemical potential is both non-zero and comparable to all other energy scales, because it is an interpolation between no-Fermi surface behavior with a single oscillation frequency $\Delta$ and two-Fermi surfaces behavior with oscillation frequencies $\Delta \pm \mu$. 

\begin{figure}
\includegraphics[width=0.4\textwidth]{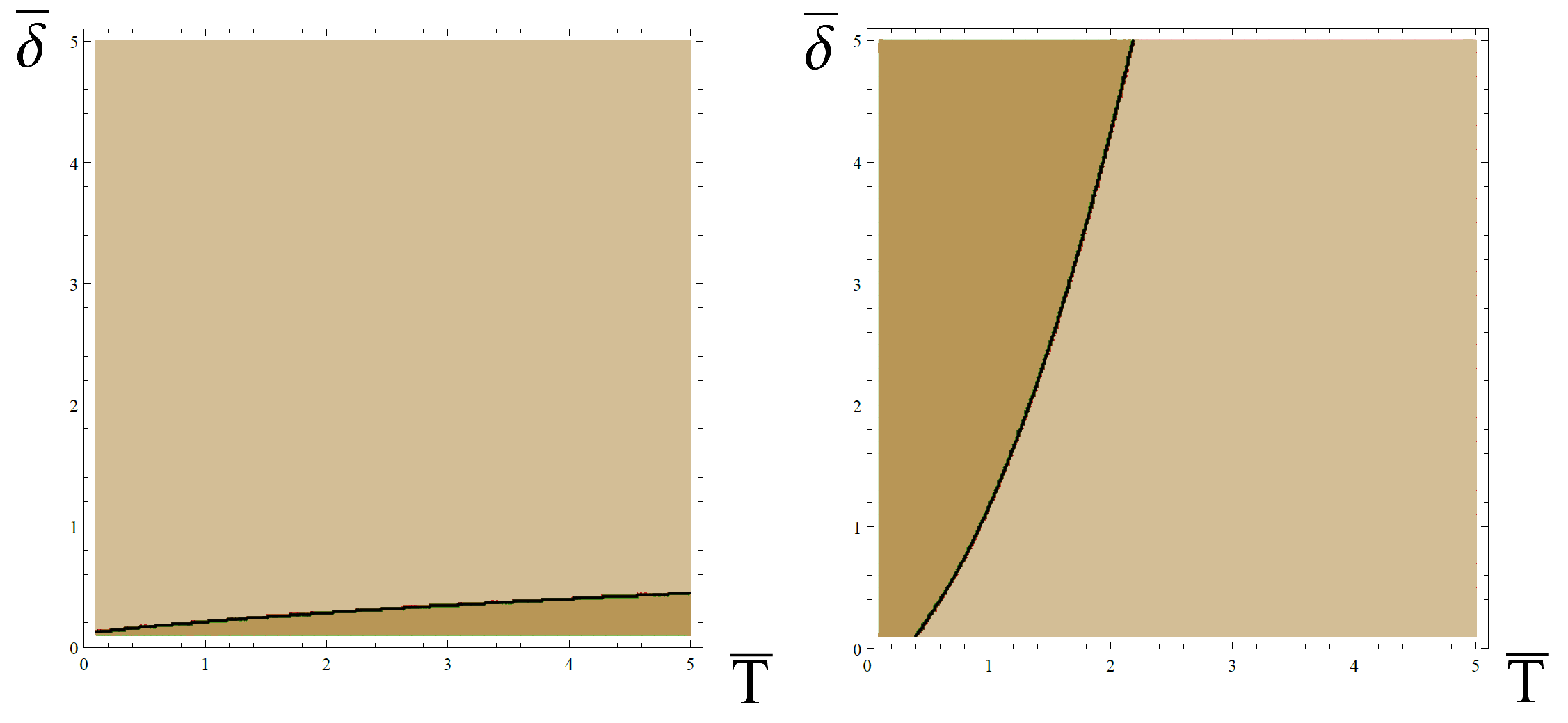}
\caption{The validities of formulas Eq.~\eqref{eq:LK-reg} (on the left) and Eq.~\eqref{eq:asympt} (on the right) as approximations to formula Eq.~\eqref{eq:oscpotmod}. The darker region represents a faithful approximation according to the $2\;\%$ criterion.}\label{fig:twoappval}
\end{figure}

So far, we have two approximations of the amplitude, high-temperature equation Eq.~\eqref{eq:LK-reg} and low-temperature equation Eq.~\eqref{eq:asympt}. The validity of these as an approximations to the full solution Eq.~\eqref{eq:oscpotmod} is shown in Fig.~\ref{fig:twoappval}. In order to gauge the faithfulness of the approximations compared to the exact result, Eq.~\eqref{eq:oscpotmod}, we arbitrarily choose $2\;\%$ relative accuracy as a criterion.

\begin{figure}
\includegraphics[width=0.4\textwidth]{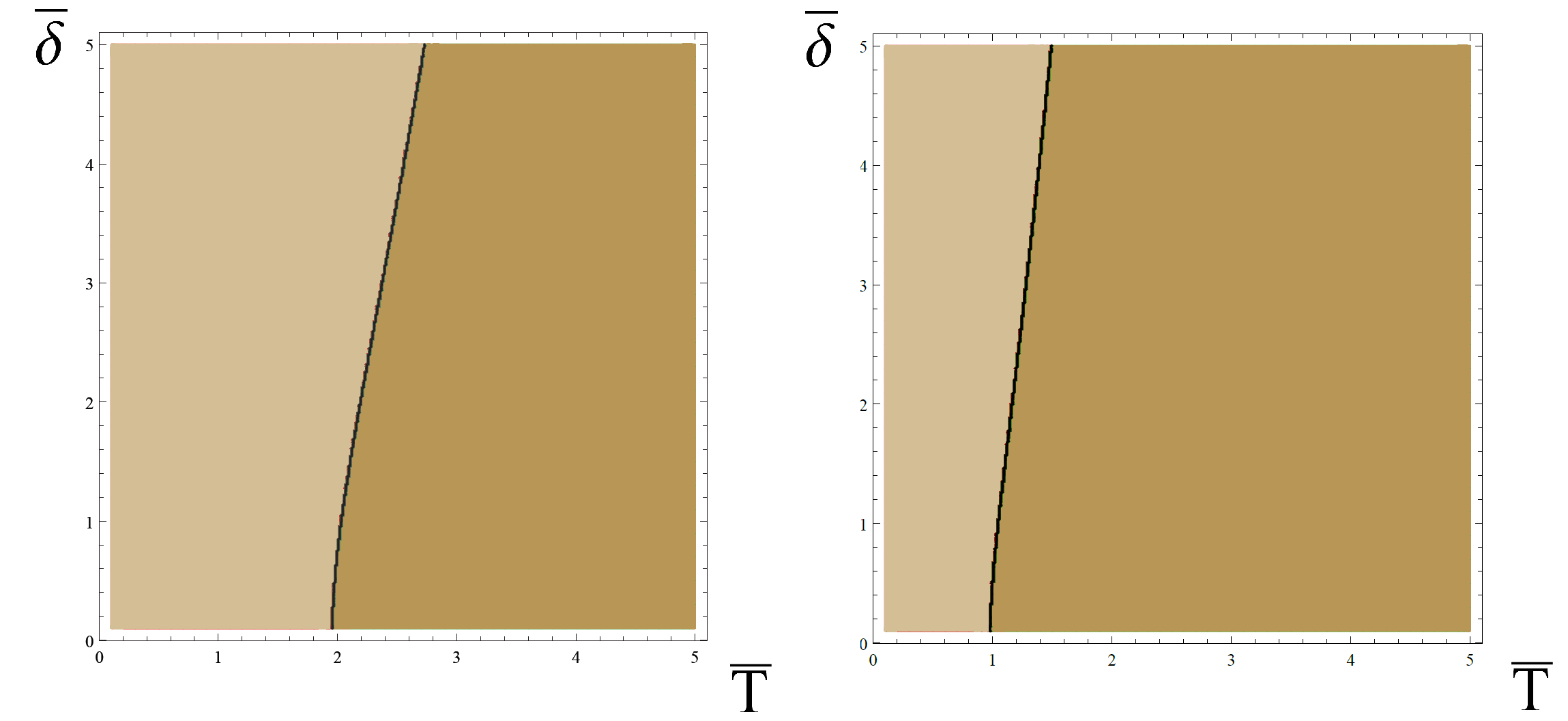}
\caption{The validities of formulas Eq.~\eqref{eq:oscpotmodtrunc0}  (on the left) and Eq.~\eqref{eq:oscpotmodtrunc1} (on the right) as approximations to formula Eq.~\eqref{eq:oscpotmod}.}\label{fig:onetwoappval}
\end{figure}

Let us discuss another type of approximation than considered above. It can be shown that if the temperature is sufficiently high, most of the contribution to the amplitude function is contained in the first few terms in Eq.~\eqref{eq:oscpotmod}. In other words, we propose the following approximation,

\begin{eqnarray}\label{eq:oscpotmodtrunc0}
\tilde{\Omega}_m^{(1)} \approx T \sum_{\omega_n>0}^{\omega_0} e^{-\frac{2 \pi l}{\omega_c} x_+}  \cos\left(\frac{2\pi l}{\omega_c}  x_{-}\right)\;,
\end{eqnarray}

\begin{eqnarray}\label{eq:oscpotmodtrunc1}
\tilde{\Omega}_m^{(2)} \approx T \sum_{\omega_n>0}^{\omega_1} e^{-\frac{2 \pi l}{\omega_c} x_+}  \cos\left(\frac{2\pi l}{\omega_c}  x_{-}\right)\;,
\end{eqnarray}
where $\tilde{\Omega}_m^{(1)}$ contains the first Matsubara mode, while $\tilde{\Omega}_m^{(2)}$ contains the first two of them. 
Due to the fact that we have increasingly suppressed exponentials in Eq.~\eqref{eq:oscpotmod}, this approximation gives a good numerical agreement in a large part of the $\overline{\delta}-\overline{T}$ plane as demonstrated in  Fig.~\ref{fig:onetwoappval}. If we combine the zero-temperature approximation and the two-term approximation, we can describe the amplitude function Eq.~\eqref{eq:oscpotmod} for all values $\overline{\delta}$ and $\overline{T}$.

\begin{figure}
\includegraphics[width=0.4\textwidth]{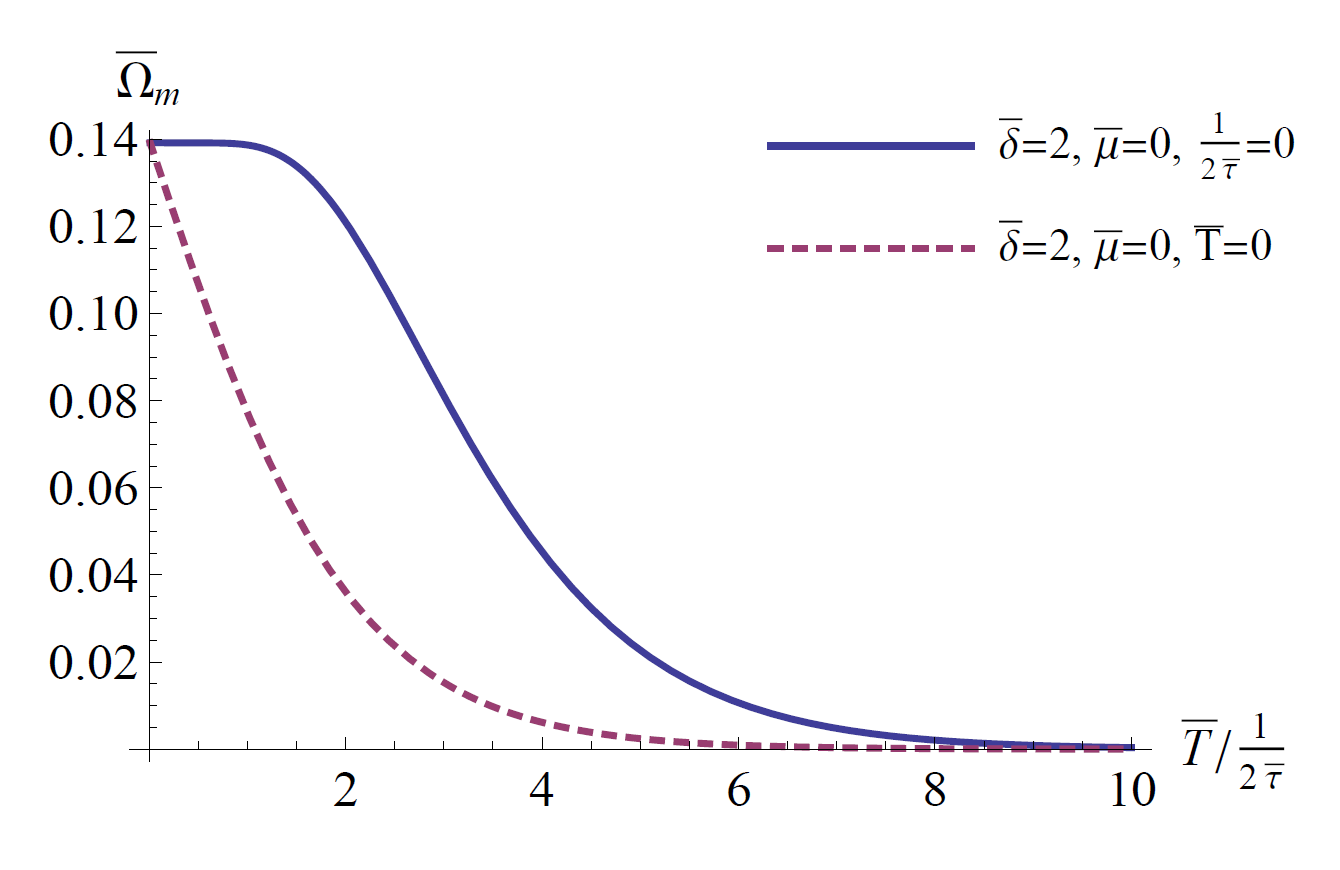}
\caption{A comparison of the amplitude function Eq.\eqref{eq:oscpotmod} as a function of temperature (setting disorder to zero) and as a function of disorder strength (setting the temperature to zero).}\label{fig:disorder}
\end{figure}

\subsection*{Analysis of the amplitude function $\tilde{\Omega}_m$, disordered case}

The presence of impurities provides an additional damping source for the amplitude function. Although it may appear that the temperature contribution is similar to the disorder contribution, they contribute in a different way, as shown in Fig.~\ref{fig:disorder} (due to more than one term being important in the summation and due to the presence of the temperature $T$ in front of the sum in Eq.~\eqref{eq:oscpotmod}).

The zero temperature result Eq.~\eqref{eq:asympt} is not valid in the presence of disorder and in this case we cannot easily calculate the amplitude function.

However, one- and two-term approximations Eq.~\eqref{eq:oscpotmodtrunc0} and Eq.~\eqref{eq:oscpotmodtrunc1} hold fairly well for finite disorder. In addition, these approximations even provide fairly accurate behavior as functions of chemical potential in the ranges of validity specified by Fig.~\ref{fig:onetwoappval}.

\section{Conclusion and Outlook}\label{sec:conclusion}
  
In this work we have presented an analytical expression for the oscillatory part of the grand potential in inverted insulators. It is valid at arbitrary magnetic field, temperature, and for weak disorder, Eq.~\eqref{eq:oscpot}. This expression can very easily be analyzed in different experimentally relevant regimes and very simple and highly accurate approximate solutions allow to determine system parameters, such as disorder; or perform an independent measurement of the gap, etc.  
As already discussed by other authors a candidate system to observe the discussed physics is bilayer graphene~\cite{Falkovsky,Alisultanov2016,McCann2006} in which the required conditions can be achieved.
A possible line of research in the future consists of investigating the effects of inelastic scattering in the present setup and see to which extent the LK formula is modified once interactions are included.

{\it {Acknowledgement:}}
 This work is also part of the D-ITP consortium, a program of the Netherlands Organisation for Scientific Research (NWO) that is funded by the Dutch Ministry of Education, Culture and Science (OCW). We acknowledge useful discussions with J. Knolle and T. Drwenski.

\end{document}